\documentstyle[12pt]{article}

\begin{document}
\begin{titlepage}

\title{Gravitational radiation of accelerated sources}

\author{J. W. Maluf$\,^{*}$, V. C. de Andrade$\,^{\S}$ and J. R. 
Steiner\\
Instituto de F\'{\i}sica, \\
Universidade de Bras\'{\i}lia\\
C. P. 04385 \\
70.919-970 Bras\'{\i}lia DF, Brazil\\}
\date{}
\maketitle

\begin{abstract}
We investigate the gravitational radiation produced by a linearly 
accelerated source in general relativity. The investigation is 
performed by studying the vacuum C metric, which is interpreted as
representing the exterior space-time of an uniformly accelerating 
spherically symmetric gravitational source, and is carried out in 
the context of the teleparallel equivalent of general relativity. 
For an observer sufficiently far from both the (modified) 
Schwarzschild and Rindler horizons, which is a 
realistic situation, we obtain a simple expression for the total 
emitted gravitational radiation. We also briefly discuss on the
absolute or relative character of the accelerated motion.
\end{abstract}
\thispagestyle{empty}
\vfill
\par
\bigskip
\noindent (*) e-mail: wadih@fis.unb.br\par
\noindent (\S) e-mail: andrade@fis.unb.br
\end{titlepage}
\newpage

\section{Introduction}
In similarity to the expected radiation of a charged particle in
classical electrodynamics, in general relativity an accelerated
source is supposed to emit gravitational radiation. Although the
magnitude of such radiation is expected to be very small, the 
understanding and description of this phenomenon is of great 
theoretical importance. The exterior space-time of an uniformly 
accelerated spherically symmetric gravitational source is described 
by the C metric, which has been the subject of renewed interest. 
The C metric is a vacuum solution of Einstein's equation first 
obtained by Levi-Civita \cite{LC}, and rediscovered subsequently by 
several authors, among whom Ehlers and Kundt \cite{EK}, who 
established its present name. Its interpretation as an accelerated 
Schwarzschild
black hole has been analyzed in Refs. \cite{KW,FZ,Bonnor}. By means
of a maximal extension of coordinates \cite{KW,Bonnor}, the C metric 
may be taken to represent a pair of accelerated black holes. As a
possible physical realization of a system that undergoes 
acceleration due to nongravitational forces, we may consider the
motion of a star (the Sun, for instance) that results from 
anisotropic solar emission.

The relevant physical property of the C metric space-time may be 
analyzed in a straightforward way by considering the linearized form 
of the solution \cite{Mashhoon}. A particle in the linearized 
C metric space-time is subject to 
the gravitational attraction of the central source plus a uniform
inertial force along the $z$ direction, say. 
Therefore the inertial state of the physical 
sources does influence the gravitational field around it, a fact 
that is consistent with the principle of equivalence. According to
the latter, gravitational and inertial forces are of the same 
nature. The constant, uniform gravitational field produced by the
accelerated source gives rise (in the linearized form of the 
solution) to the gravitational Stark effect \cite{Mashhoon}.

In this article we investigate the gravitational radiation 
emitted by an accelerated Schwarzschild black hole in the 
framework of the teleparallel equivalent of general relativity 
(TEGR). In our investigation we will adopt a Bondi type coordinate
system that allows us to conclude that the full, nonlinearized 
form of the C metric may be understood as a nonlinear
superposition of the Rindler and Schwarzschild space-times 
\cite{Mashhoon}. The C metric exhibits a timelike Killing vector 
$\partial_t$, which indicates the static character of the metric. 
However, the metric is also radiative. As discussed in Ref. 
\cite{KW}, this apparent paradox is dispelled by noting that the 
Killing vector above becomes spacelike beyond the Rindler horizon.
Therefore the metric is not globally static. Morevover, the 
Riemann tensor contains $r^{-1}$ terms, which indicate the
radiative character of the metric. 

In the description of the C
metric given by Ref. \cite{Mashhoon} the coordinate system is
adapted to (centered on) the accelerated source. Therefore in
our analysis we will perform a time dependent Lorentz boost in 
the opposite direction of the moving black hole that will 
determine a set of tetrad fields adapted to observers that are
``approximately at rest" in space-time. In other words, with 
respect to the boosted set of tetrad fields the black hole does 
indeed display an accelerated motion, in an approximation to be
explained. The energy and the gravitational radiation emitted by the 
black hole will be evaluated with respect to this set of observers. 
The surface integrals will be evaluated on a surface of integration 
that is simultaneously far from both (Schwarzschild and Rindler) 
modified horizons.

The article is organized as follows. In section II we present the C
metric in Bondi type coordinates, and briefly discuss the relevant
property of the linearized form of the metric. In section III we 
recall some relevant definitions that arise in the TEGR 
regarding the energy-momentum and the flux of energy-momentum of the
gravitational field. In section IV we carry out, first, the 
construction of a set of (time independent) tetrad fields adapted
to the accelerated black hole (i.e., for which the black hole is at
rest). Then by means of a local Lorentz
transformation we determine the tetrad fields which describe the
accelerated motion of the source. After presenting the 
relevant calculations and approximations we arrive at a
simple expression for the total gravitational radiation flux, which 
is the major result of the article. In section V we discuss on the 
relativity of acceleration. For this purpose, we examine whether we 
may attribute to the accelerated motion of the black hole in 
space-time a relative or absolute character.

Notation: space-time indices $\mu, \nu, ...$ and SO(3,1)
indices $a, b, ...$ run from 0 to 3. Time and space indices are
indicated according to
$\mu=0,i,\;\;a=(0),(i)$. The tetrad field $e^a\,_\mu$ 
yields the definition of the torsion tensor:  
$T^a\,_{\mu \nu}=\partial_\mu e^a\,_\nu-\partial_\nu e^a\,_\mu$.
The flat, Minkowski space-time  metric is fixed by
$\eta_{ab}=e_{a\mu} e_{b\nu}g^{\mu\nu}= (-+++)$.        

\section{The C metric}

Except for a change in signature, we will adopt the notation of 
Ref. \cite{Mashhoon} in the description of the C metric.
By means of a coordinate transformation the 
C metric in the form given by Ehlers and Kundt \cite{EK} may be 
written in Bondi type coordinates $(u,r,\theta,\phi)$ (which may
be understood as an accelerated coordinate system)  in terms of 
two functions, $G(\theta)$ and $H(r,\theta)$ \cite{Mashhoon}. We
find it useful to define the function $g(\theta)$ according to
$G=g^2\sin^2\theta$. In terms of this function the C metric is
written as

\begin{equation}
ds^2=-H\,du^2-2\,du\,dr+2Ar^2\,\sin\theta\,dud\theta+
{r^2\over g^2}\,d\theta^2+r^2g^2\,\sin\theta^2\,d\phi^2\,,
\label{1}
\end{equation}
where

\begin{equation}
g^2\sin^2\theta = G(\theta) =
\sin^2\theta-2mA\,\cos^3\theta\,,
\label{2}
\end{equation}

\begin{eqnarray}
H(r,\theta)&=&1-{{2m}\over r} -A^2r^2(\sin^2\theta-2mA\cos^3\theta)
\nonumber \\
&-&2Ar\cos\theta(1+3mA\cos\theta) + 6mA\cos\theta\,.
\label{3}
\end{eqnarray}
The parameters $m>0$ and $A>0$ represent the mass and acceleration 
of the black hole, respectively. It is not difficult to show that 
$G>0$ provided the relation $mA<1/(3\sqrt{3})$ is satisfied. We will
assume this relation to hold in the present analysis.

In the C metric space-time there are two horizons, the Schwarzshild 
and Rindler horizons located at $r_S$ and $r_R$, respectively. Let
us define the acceleration length $L_A=1/(3\sqrt{3}A)$, and the 
functions

$$U=-{1\over \sqrt{3}} \cos\biggl(
{1\over 3} \arccos {m\over {L_A}}\biggr)\,,$$

$$V={1\over \sqrt{3}}\sin\biggl(
{1\over 3}\arccos{m\over {L_A}}\biggr)\,.$$
In terms of these quantities, $r_S$ and $r_R$ are given by
\cite{Bini}

$$r_S={1\over A} {{\sqrt{3}V-U} \over
{\lbrack1+ (\sqrt{3} V -U)\cos\theta \rbrack}}\,,$$

$$r_R={1\over A} {{2U}\over {(1+2U\cos\theta)}}\,.$$
In the limit $mA<<1$ the quantities above reduce to \cite{FZ2}

$$r_S\cong 2m(1+2Am\cos\theta)\,,$$

$$r_R\cong {1\over {
A(1-\cos\theta+Am\sin^2\theta)}}\,.$$

The C metric may be interpreted as a nonlinear suporposition of the 
Schwarzschild and Rindler space-times. This interpretation may be 
verified by investigating the limits of vanishing parameters. By 
making $A=0$ the metric tensor above reduces to

\begin{equation}
ds^2=-\biggl(1-{{2m}\over r}\biggr)du^2-2\,du\,dr
+r^2\,d\theta^2+r^2\,\sin\theta^2\,d\phi^2\,,
\label{4}
\end{equation}
This is precisely the Schwarzschild metric tensor written in terms
of the retarded time $u=t-r-2m\ln (r/2m\,-1)$. 

On the other hand, by requiring $m=0$ we obtain

\begin{eqnarray}
ds^2&=&-(1- 2Ar\cos\theta - A^2r^2\sin^2\theta )du^2
-2\,du\,dr \nonumber \\
&+&2Ar^2\sin\theta du d\theta
+r^2\,d\theta^2+r^2\,\sin\theta^2\,d\phi^2\,,
\label{5}
\end{eqnarray}
This metric tensor can be transformed into Minkowski's metric tensor
by means of the coordinate transformation \cite{KW}

\begin{eqnarray}
\bar{t}&=&(A^{-1}-r\cos\theta) \sinh Au + r\cosh Au\,, \nonumber \\
\bar{z}&=&(A^{-1}-r\cos\theta) \cosh Au + r\sinh Au\,, \nonumber \\
\bar{x}&=&r\sin\theta\cos\phi\,, \nonumber \\
\bar{y}&=&r\sin\theta\sin\phi\,.
\label{6}
\end{eqnarray}
The coordinate transformation above transforms Eq. (5) into

\begin{equation}
ds^2=-d\bar{t}^2+d\bar{x}^2+d\bar{y}^2+d\bar{z}^2\,.
\label{7}
\end{equation}
In the space-time represented by the coordinates 
$(\bar{t},\bar{x},\bar{y},\bar{z})$ the locus $r=0$ determines a
timelike curve given by

\begin{eqnarray}
\bar{t}&=&A^{-1}\sinh Au\,, \nonumber \\
\bar{z}&=&A^{-1}\cosh Au \,, \nonumber \\
\bar{x}&=&\bar{y}=0\,,
\label{8}
\end{eqnarray}
which represents one branch of the hyperbola 
$\bar{z}^2-\bar{t}^2=1/A^2$, along which takes place the motion
with constant acceleration $A$, parametrized in terms of $u$. The
Bondi type coordinate system $(u,r,\theta,\phi)$ 
covers only the half-space $\bar{t}+\bar{z}>0$. The maximal 
extension consists in filling the remaining half-space by replicating 
a time-reversed copy of the half-space $\bar{t}+\bar{z}>0$. In this
case the hyperbola $\bar{z}^2-\bar{t}^2=1/A^2$ represents two 
uniformly accelerated particles, moving in opposite directions, and 
which are causally disconnected, because there is no overlaping of 
their fields in space-time. For further details, see Ref. \cite{KW}.

The linearized form of the C metric is simply obtained by
neglecting $m^2$, $mA$, $A^2$ and higher-order terms
\cite{Mashhoon}. By 
transforming $(u,r,\theta,\phi)$ to $(T,X,Y,Z)$ coordinates
according to 

\begin{eqnarray}
T&=&u+r+2m\ln\biggl( {r\over {2m}}-1\biggr) \nonumber \\
X&=&r\sin\theta\cos\phi \nonumber \\
Y&=&r\sin\theta\sin\phi \nonumber \\
Z&=&r\cos\theta\,,
\label{9}
\end{eqnarray}
we find that in the linearized limit the $g_{00}$ component of 
the metric tensor given by Eq. (1) reads
 
\begin{eqnarray}
-g_{00}(u,r,\theta,\phi)&=&-g_{00}(T,X,Y,Z) \nonumber \\
& \approx & 1-{{2m}\over r} - 2Ar\cos\theta\,.
\label{10}
\end{eqnarray}
The $g_{00}$ component allows us to identify the Newtonian 
potential $\phi$ according to

\begin{equation}
-g_{00}(T,X,Y,Z)=1+2\phi\,.
\label{11}
\end{equation}
We obtain

\begin{equation}
\phi=-{m\over r}-(r\cos\theta) A = -{m\over r}-AZ\,.
\label{12}
\end{equation}
The potential $\phi$ determines the Newtonian equation of 
motion for a point particle in this space-time,

\begin{equation}
{{d^2X^i}\over{dT^2}}= 
-{{\partial \phi}\over {\partial X^i}}\,.
\label{13}
\end{equation}
from what follows

\begin{equation}
{{d^2 {\bf r}}\over{dT^2}}= 
-{m\over r^3} {\bf r}+A\hat{{\bf z}} \,.
\label{14}
\end{equation}
Therefore in the linearized C metric space-time a point particle is
suject to the usual central force plus an additional constant and 
uniform force that is due to the accelerated motion of the source,
represented by $m$, along the $\theta=\pi$ direction.

In what follows we will need the inverse metric tensor associated 
to Eq. (1). It reads

\begin{equation}
g^{\mu\nu}=\pmatrix{ 0&-1&0&0 \cr
-1& H+A^2r^2g^2\sin^2\theta & A g^2 \sin\theta&0 \cr
0&A g^2\sin\theta& {g^2\over r^2}& 0 \cr
0&0&0&{1\over{r^2g^2\sin^2\theta}}\cr}\,.
\label{15}
\end{equation}

\section{The TEGR and the gravitational energy-momentum}

The TEGR is a reformulation of Einstein's general relativity in 
terms of the tetrad field $e^a\,_\mu$
\cite{Moller,Hay,Hehl,Nester,Maluf1,Pereira,Obukhov}. 
By considering the theory to be invariant under the global SO(3,1) 
group of transformations of $e^a\,_\mu$, the additional
six degrees of freedom of the latter, with respect to the metric
tensor $g_{\mu\nu}$, fix the reference frame adapted to particular
observers.

The Lagrangian density for the gravitational field in the TEGR, in
the presence of matter fields, is given by

\begin{eqnarray}
L(e_{a\mu})&=& -k\,e\,({1\over 4}T^{abc}T_{abc}+
{1\over 2} T^{abc}T_{bac} -T^aT_a) -L_m\nonumber \\
&\equiv&-k\,e \Sigma^{abc}T_{abc}-L_m\;,
\label{16}
\end{eqnarray}
where $k=1/(16 \pi G)$ and $e=\det(e^a\,_\mu)$. The
tensor $\Sigma^{abc}$ is defined by

\begin{equation}
\Sigma^{abc}={1\over 4} (T^{abc}+T^{bac}-T^{cab})
+{1\over 2}( \eta^{ac}T^b-\eta^{ab}T^c)\;,
\label{17}
\end{equation}
and $T^a=T^b\,_b\,^a$. The quadratic combination
$\Sigma^{abc}T_{abc}$ is proportional to the scalar curvature
$R(e)$, except for a total divergence \cite{Maluf1}. $L_m$
represents the Lagrangian density for matter fields. The field
equations for the tetrad field read

\begin{equation}
e_{a\lambda}e_{b\mu}\partial_\nu(e\Sigma^{b\lambda \nu})-
e(\Sigma^{b \nu}\,_aT_{b\nu \mu}-
{1\over 4}e_{a\mu}T_{bcd}\Sigma^{bcd})
\;=\;{1\over {4k}}eT_{a\mu} \;,
\label{18}
\end{equation}
where $\delta L_m/\delta e^{a\mu}\equiv eT_{a\mu}$
It is possible to prove by explicit calculations that the left hand
side of Eq. (18) is exactly given by ${1\over 2}\,e\,
\lbrack R_{a\mu}(e)-{1\over 2}e_{a\mu}R(e)\rbrack$.

The definition of the gravitational energy-momentum contained within
an arbitrary volume $V$ of the three-dimensional spacelike
hypersurface arises in the Hamiltonian formulation of the
TEGR \cite{Maluf2}. It reads

\begin{equation}
P^a=-\int_V d^3x\, \partial_j \Pi^{aj}\;,
\label{19}
\end{equation}
where $\Pi^{aj}=-4ke\Sigma^{a0j}$ is the momentum canonically
conjugated to $e_{aj}$.  
An essential feature of the gravitational energy-momentum
$P^a=(E,{\bf P})$ is the covariance under  global SO(3,1) 
transformations, in addition to the invariance
under coordinate transformations of the three-dimensional
spacelike hypersurface. Each configuration of tetrad fields
establishes a reference frame adapted to an observer (see the
discussion at the beginning of the following section).

Simple algebraic manipulations of Eq. (18)
yield a continuity equation for the gravitational energy-momentum
$P^a$ \cite{Maluf3,Maluf6,Maluf4},

\begin{equation}
{{d P^a} \over {dt}} =
-\Phi^a_g -\Phi^a_m\;,
\label{20}
\end{equation}
where

\begin{equation}
\Phi^a_g=k\int_S dS_j\lbrack e e^{a\mu}
(4\Sigma^{bcj}T_{bc\mu}-\delta^j_\mu \Sigma^{bcd}T_{bcd})\rbrack\;,
\label{21}
\end{equation}
is $a$ component of the  gravitational energy-momentum flux, and

\begin{equation}
\Phi^a_m=\int_S dS_j\,(ee^a\,_\mu T^{j\mu})\,,
\label{22}
\end{equation}
is the $a$ component of the matter energy-momentum flux. $S$
represents the spatial boundary of the volume $V$. Therefore in the
vacuum space-time the gravitational energy flux $\Phi^{(0)}_g$
is simply given by

\begin{equation}
\Phi^{(0)}_g=-{{dE}\over {dt}} \;,
\label{23}
\end{equation}
where $E=P^{(0)}$. The analysis \cite{Maluf3,Maluf6} of some 
relevant and well known configurations of the gravitational field 
has supported the interpretation of $\Phi^{(0)}_g$ as the 
gravitational energy flux.

\section{Reference frames in the C metric space-time and the 
gravitational radiation}

A given gravitational field configuration described by the metric
tensor $g_{\mu\nu}$ admits an infinity of tetrad fields, related to
each other by means of a local SO(3,1) transformation. In order to 
examine how an observer is adapted to a particular set of 
tetrad fields, we consider its worldline in space-time. 
Let $x^\mu(s)$ denote the worldline $C$ of an observer,
and $u^\mu(s)=dx^\mu/ds$ its velocity along $C$. We may identify
the observer's velocity with the $a=(0)$ component of $e_a\,^\mu$,
where $e_a\,^\mu e^a\,_\nu=\delta^\mu_\nu$. Thus, 
$u^\mu(s)=e_{(0)}\,^\mu$ along $C$ \cite{Hehl1}. 
The acceleration of the observer is given by 

\begin{equation}
a^\mu= {{Du^\mu}\over{ds}}={{De_{(0)}\,^\mu }\over{ds}}=
u^\alpha \nabla_\alpha e_{(0)}\,^\mu\,.
\label{24}
\end{equation}

The covariant derivative is constructed out of the Christoffel 
symbols. We see that $e_a\,^\mu$
determines the velocity and acceleration along a worldline of an 
observer adapted to the frame.
From this perspective we conclude that a given set of 
tetrad fields, for which $e_{(0)}\,^\mu$ describes a congruence of
timelike curves, is adapted to a particular class of observers,
namely, to observers determined by the velocity field 
$u^\mu=e_{(0)}\,^\mu$, endowed with acceleration $a^\mu$. 
In the case of asymptotically flat space-times, for instance,
if $e^a\,_\mu \rightarrow \delta^a_\mu$ in the asymptotic limit 
$r \rightarrow \infty$, then we conclude that $e^a\,_\mu$ is 
adapted to stationary observers at spacelike infinity. 

Let us now consider the C metric. In order to determine 
$e^a\,_\mu$ we will impose restrictions on 
$e_{(0)}\,^\mu=g^{\mu\lambda} e_{(0)\lambda}$. With the help
of Eq. (15) we require 

\begin{equation}
e_{(0)}\,^1=e_{(0)}\,^2=e_{(0)}\,^3=0\,.
\label{25}
\end{equation}
The requirement of the equations above establishes a set
of tetrad fields adapted to the moving black hole, i.e., in such
reference frame the observer sees the black hole at rest.

It is not difficult to implement Eq. (25) in the construction of
the tetrad field $e^a\,_\mu$ that yields the C metric given by 
Eq. (1). After few algebraic manipulations we obtain

$$e^a\,_\mu(u,r,\theta,\phi)=$$

\begin{equation}
=\pmatrix{
B&C&-D&0\cr
0&C\sin\theta\cos\phi&
{r\over g}\cos\theta\cos\phi-D\sin\theta\cos\phi&
-rg\,\sin\theta\sin\phi \cr
0&C\sin\theta\sin\phi&
{r\over g}\cos\theta\sin\phi-D\sin\theta\sin\phi&
rg\,\sin\theta\cos\phi \cr
0&C\cos\theta&-{r\over g} \sin\theta-D\cos\theta&0}
\label{26}
\end{equation}
where $g$ is given by Eq. (2) and 

\begin{eqnarray}
B&=&H^{1/2} \nonumber \\
C&=&H^{-1/2} \nonumber \\
D&=&H^{-1/2}Ar^2\sin\theta 
\label{27}
\end{eqnarray}
The tetrad field above satisfies Eq. (25), and therefore it is 
adapted to the accelerated black hole. We further note that by
requiring $m=A=0$ the tetrad field above reduces to

$$e^a\,_\mu(u,r,\theta,\phi)=\pmatrix{
1&1&0&0\cr
0&\sin\theta \cos\phi&r\cos\theta \cos\phi&-r\sin\theta\sin\phi \cr
0&\sin\theta \sin\phi&r\cos\theta \sin\phi&r\sin\theta\cos\phi \cr
0&\cos\theta&-r\sin\theta&0}\,,
$$
which is the flat space-time tetrad field in Bondi type coordinates.
All components of the torsion tensor
$T^a\,_{\mu \nu}=\partial_\mu e^a\,_\nu-\partial_\nu e^a\,_\mu$ 
calculated out of the expression above vanish.

Given that Eq. (26) is time independent, we conclude that the 
resulting expression for $P^{(0)}$
is likewise time independent. Therefore in view of Eq. (23) we have
$\Phi^{(0)}_g=0$. Thus from the standpoint of the set of observers 
for whom the black hole is at rest, as determined by Eq. (26), 
there is no gravitational radiation. 

In the analysis of the linearized form of the C metric in section 2
we observed that the equation of motion given by Eq. (14), for a
free point particle in the C metric space-time, is consistent with 
the picture of an accelerated black hole in the $-z$ direction.
Therefore we will consider a boost transformation $\Lambda^a\,_b$
(a {\it local} Lorentz transformation)
along the $+z$ direction, and apply it on $e^a\,_\mu$ given by Eq. 
(26). The boost is given by

\begin{equation}
\Lambda^a\,_b=\pmatrix{
\gamma&0&0&-\gamma v\cr
0&1&0&0\cr
0&0&1&0\cr
-\gamma v&0&0&\gamma}\,,
\label{28}
\end{equation}
where $v=v(u)$ and $\gamma=(1-v^2)^{-1/2}$. The resulting tetrad
field $\tilde{e}^a\,_\mu = \Lambda^a\,_b e^b\,_\mu$ reads 
(dropping the tilde)

$$e^a\,_\mu(u,r,\theta,\phi)=$$

\begin{equation}
=\pmatrix{
\gamma B&\gamma C -\gamma v C\cos\theta&
-\gamma D -\gamma v(-{r\over g} \sin\theta-D\cos\theta) &0\cr
0&C\sin\theta\cos\phi&
{r\over g}\cos\theta\cos\phi-D\sin\theta\cos\phi&
-rg\,\sin\theta\sin\phi \cr
0&C\sin\theta\sin\phi&
{r\over g}\cos\theta\sin\phi-D\sin\theta\sin\phi&
-rg\,\sin\theta\cos\phi \cr
-\gamma vB&-\gamma vC+\gamma C\cos\theta&
\gamma vD +\gamma(-{r\over g} \sin\theta-D\cos\theta)&0}
\label{29}
\end{equation}
We remark that the tetrad field above yields the complete C metric 
tensor, not its linearized form.

In the remaining of this section we will evaluate $T_{a\mu\nu}$ and
$\Sigma^{\lambda\mu\nu}$ out
of Eq. (29). Let us note, however, that 
the tetrad field given by the equation above is adapted to the set
of observers whose four-velocity field 
$e_{(0)}\,^\mu=g^{\mu\lambda} e_{(0)\lambda}$ in  
$(u,r,\theta,\phi)$ coordinates reads

\begin{equation}
e_{(0)}\,^\mu=
\biggl( \gamma C(1-v\cos\theta), \gamma v(B\cos\theta-
Arg\sin\theta),
-\gamma v {g\over r}\sin\theta,0\biggr)\,.
\label{30}
\end{equation}

In order to carry out the calculations we will establish two 
essential approximations. We will restrict considerations to the
space-time region for which

\begin{eqnarray}
{m\over r}& << &1 \\
Ar &<<&1
\label{31,32}
\end{eqnarray}
Eqs. (31) and (32) ensure that an observer located at the radial
position $r$ is far from the Schwarzschild and Rindler horizons,
respectively, and may be taken as conditions for a weak field 
approximation. Let us consider a small parameter $\varepsilon$
and stipulate that 

\begin{eqnarray}
{m\over r}& \cong& O(\varepsilon) \\ \nonumber
Ar &\cong & O(\varepsilon)\,.
\label{33}
\end{eqnarray}
It follows that $(m/r)Ar=mA=O(\varepsilon^2)$. In this approximation
we have $g =1+O(\varepsilon^2)$, $B=1+O(\varepsilon)$ and also
$C=1+O(\varepsilon)$. It is easy to verify that in the approximation
determined by Eqs. (31,32) the four-velocity given by Eq. (30)
reduces to

\begin{equation}
e_{(0)}\,^\mu \cong (\gamma, 0, 0, \gamma v)\,,
\label{34}
\end{equation}
in $(T,X,Y,Z)$ coordinates, which we recognize as the usual 
four-velocity of observers boosted in the $+z$ direction in flat
space-time. 

Therefore in the approximation determined by Eqs.
(31,32) the Lorentz boost given by Eq. (28) corresponds to an
ordinary boost in flat space-time. Since both $(u,r,\theta,\phi)$
and $(T,X,Y,Z)$ are accelerated coordinate systems, Eq. (34)
establish the four-velocity of observers that are approximately at
rest in space-time.

We return to Eq. (29) and proceed to evaluate the necessary 
quantities that lead to 

\begin{equation}
P^{(0)}=-\int_V d^3x\, \partial_j \Pi^{(0)j}=
4k\oint_S d\theta d\phi\, e\Sigma^{(0)01}\;,
\label{35}
\end{equation}
according to  Eq. (19). In the expression above $S$ is a surface of
constant radius $r$ that satisfies Eqs. (31,32),
$e=\det (e^a\,_\mu)=r^2\sin\theta$, and

$$\Sigma^{(0)01}=
e^{(0)}\,_0\Sigma^{001}+e^{(0)}\,_1\Sigma^{101}+
e^{(0)}\,_2\Sigma^{201}\,.$$
Equation (35) determines the gravitational energy contained within
a large surface of constant radius $r$, which is also the radial
position of an observer. It is not the total gravitational energy
of the space-time, but since the surface $S$ encloses the black
hole, it allows the determination of the gravitational radiation,
as we will see. In view of the fact that the radius $r$ satisfies 
Eqs. (31,32), the surface of integration $S$ is far from both the 
Schwarzschild and Rindler horizons.

We have the following nonvanishing, exact expressions for 
$T_{\lambda\mu\nu}=e^a\,_\lambda T_{a\mu\nu}$:

\begin{eqnarray}
T_{012}&=&B(\partial_r D + \partial_\theta C)\,, \\ \nonumber
T_{102}&=& C\partial_\theta B-\gamma^2\dot{v}\,C{r\over g}\sin\theta
\,,\\ \nonumber
T_{201}&=& -D\partial_r B +\gamma^2\dot{v}\,{r\over g}\, C\sin\theta
\,,\\ \nonumber
T_{202}&=& D\partial_\theta B\,, \\ \nonumber
T_{212}&=&{r\over g}\biggl( {1\over g} -C\biggr)\,,\\ \nonumber
T_{313}&=& rg(g-C)\sin^2\theta \,,\\ \nonumber
T_{323}&=&-(1-g^2)r^2\sin\theta \cos\theta-
rg(D-r\partial_\theta g)\sin^2\theta \,.
\label{36}
\end{eqnarray}
where $\dot{v}=dv/du$, and the functions $B$, $C$ and
$D$ are defined by Eq. (27).
Taking in account definition (17) we obtain

\begin{eqnarray}
\Sigma^{001}&=& -{1\over {2r}}\biggl[ 2-
\biggl(g+{1\over g}\biggr)C\biggr], \\ \nonumber
\Sigma^{101}&=& {1\over 2}\biggr[ {g^2\over r^2}\,D\partial_\theta B
+(1-g^2)A\cos\theta , \\ \nonumber
&\;&+A\,{g\over r}\,(D-r\partial_\theta g)\sin\theta\biggr] ,
\\ \nonumber
\Sigma^{201}&=& {g^2 \over{2r^2}}\,B\partial_\theta C+
{g^2 \over {4r^2}}\,\partial_r(BD) +{g\over {2r}}\,A(g-C)\sin\theta .
\label{37}
\end{eqnarray}

The exact expression for the integrand in Eq. (35) is given by

\begin{eqnarray}
e\Sigma^{(0)01}&=&
\gamma B \sin\theta\biggl\{-{r\over 2}\biggl[2-
\biggl( g+{1\over g} \biggr)C\biggr] \biggr\} \\ \nonumber
&\;&+(\gamma C-
\gamma v\,C\cos\theta)\sin\theta\biggr[{1\over 2}\,g^2
D\partial_\theta B\\ \nonumber
&\;&+{1\over 2}(1-g^2)Ar^2\cos\theta+
{1\over 2}Ar\,g(D-r\partial_\theta g)\sin\theta \biggr]\\ \nonumber
&\;&+(-\gamma D+ \gamma v\,{r\over g}\,\sin\theta+
\gamma v D\cos\theta)\sin\theta
\biggl[{1\over 2} g^2\,B\partial_\theta C \\ \nonumber
&\;&+{1\over 4}g^2\partial_r (BD)+
{1\over 2}Ar\,g(g-C)\sin\theta \biggr]\,.
\label{38}
\end{eqnarray}

We proceed now to simplify the expression above by taking into
account Eq. (33). Specifically, we wish to write it up to the first
power in $\varepsilon$. We note first that $g=1+O(\varepsilon^2)$ 
and also $g^{-1}=1+O(\varepsilon^2)$. Thus in the first order
approximation we may take $g\cong 1$, and therefore 
$e\Sigma^{(0)01}$ is simplified to

\begin{eqnarray}
e\Sigma^{(0)01}&\cong&
-\gamma \, r(B-1) \sin\theta \\ \nonumber
&\;&+{1\over 2}\gamma(C-
v\,C\cos\theta)\sin\theta( 
D\partial_\theta B + 
Ar D\sin\theta) \\ \nonumber
&\;&+{1\over 2}\gamma(-D+ v\,r\,\sin\theta+
v D\cos\theta)\sin\theta
\biggl[-C\partial_\theta B \\ \nonumber
&\;&+{1\over 2}\partial_r (BD)+
Ar\,(1-C)\sin\theta \biggr]\,.
\label{39}
\end{eqnarray}
Next we expand the functions $B$ and $C$ up to $O(\varepsilon^2)$.
We find

\begin{eqnarray}
B&\cong& 1-{m\over r}-\biggl(Ar+{1\over 2}mA\biggr)\cos\theta-
{1\over 2}\biggl({m\over r}\biggr)^2\\ \nonumber
&\;&-{1\over 2}(Ar)^2\sin^2\theta-{1\over 8}(Ar)^2\cos^2\theta\,,
\\ \nonumber
C&\cong& 1+{m\over r}+\biggl(Ar+{1\over 2}mA\biggr)\cos\theta+
{1\over 2}\biggl({m\over r}\biggr)^2\\ \nonumber
&\;&+{1\over 2}(Ar)^2\sin^2\theta+{1\over 8}(Ar)^2\cos^2\theta\,.
\label{40}
\end{eqnarray}
We observe that

\begin{eqnarray}
B-1&=&-{m\over r}-Ar\cos\theta+O(\varepsilon^2)\,,\\ \nonumber
\partial_\theta B&=& (Ar)\sin\theta+O(\varepsilon^2)\,.
\label{41}
\end{eqnarray}
Upon substitution of Eqs. (40,41) into Eq. (39) we obtain

\begin{equation}
e\Sigma^{(0)01}=\gamma m \sin\theta +
\gamma(Ar)\sin\theta \cos\theta +\gamma O(\varepsilon^2)\,.
\label{42}
\end{equation}
Given that the integral of the second term in the expression above
vanishes upon integration in $\theta$ between $0$ and $\pi$, we
finally integrate Eq. (35) and obtain 

\begin{equation}
P^{(0)} \cong {1\over {4\pi}}
\oint_S d\theta d\phi\,( \gamma\,m\sin\theta)
= \gamma\,m\,.
\label{43}
\end{equation}

The expression above corresponds to the the gravitational 
energy evaluated with respect to observers whose four-velocity
is given by Eq. (30), which in first approximation reduces to
Eq. (34).  For such observers, the closed surface of integration 
$S$ in Eq. (35) is sufficiently far from the black hole and from
the Rindler horizon. 

The total flux of gravitational radiation is determined by Eq. 
(23). We easily find

\begin{equation}
\Phi^{(0)}_g = \dot{P}^{(0)}\cong mv\dot{v}\gamma^3\,.
\label{44}
\end{equation}
For an uniformly accelerated motion we identify

$$\dot{v}={1\over \gamma^3} A\,,$$
This identification is normally considered in the context of an 
uniformly accelerated particle in special relativity (see, for
instance, Ref. \cite{Inv}). Therefore we finally obtain for the 
total gravitational radiation flux

\begin{equation}
\Phi^{(0)}_g \cong mvA\,.
\label{45}
\end{equation}
Had we considered higher order terms in Eqs. (39-42), we would obtain
additional terms of the type $O(\varepsilon^3)$ in the expression 
above. This fact can be verified by noting that the time derivative
of $\gamma$ in the last term of Eq. (42), 
$\gamma O(\varepsilon^2)$, yields a term of the type
$A\,O(\varepsilon^2)=(1/r)(Ar)O(\varepsilon^2)=(1/r)O(\varepsilon^3)$,
which is negligible for large values of $r$.

We remark that by making $v \rightarrow -v$ we would be considering a
totally different physical configuration, that has no particular 
relevance to our analysis. However, in view of Eq. (44), we note that
$\Phi^{(0)}_g$ would remain positive.

The gravitational radiation emitted by an uniformly accelerated
source, as described by the C metric, has been previously 
investigated by Farhoosh and Zimmerman \cite{FZ}, and by
Tomimatsu \cite{AT}. The idea of the approach by these
authors is to bring the C metric
into the Bondi-Metzner-Sachs form \cite{BMS} and to take the time
derivative of the mass aspect, that yields minus the square of 
news function. In this case the mass loss and the emitted radiation
can be easily obtained. However the C metric does not display the 
asymptotic boundary conditions of the Bondi radiative solution, 
and therefore this scheme works only in a certain approximation 
(in Bondi's space-time there is no Rindler horizon at large radial
distances). Farhoosh and Zimmerman address the C metric 
gravitational radiation by  constructing an {\it inertial}
coordinate system out of the {\it accelerated} one.
They conclude that to lowest order in acceleration, the
mass loss is proportional to $(Am)^2$, in contrast to our result
given by Eqs. (44,45), that is linear in $Am$. We note, however, 
that we work with a unique coordinate system. 

A possible small acceleration of the Sun has been considered 
recently as an explanation of the Pioneer anomaly
\cite{Anderson,MMN}.
An anomalous acceleration on the motion of the Pioneer 10 and 
Pioneer 11 spacecrafts has been measured, which is directed towards
the Sun. The suggested explanation \cite{Mashhoon} is based on a 
slight modification of Eq. (14) for the spacecrafts, possibly due 
to a nongravitational force exerted on the Sun as a consequence of
anisotropic solar emission. We note that 
the approximations given by Eqs. (31,32) are quite natural in this
physical scenario. An observer on Earth located at the radial
position $r$ from the center of the Sun is very distant from the 
Rindler horizon. The gravitational radiation emitted by the Sun 
and measured on Earth is evaluated through a solid angle of 
constant radius $r$.

\section{The absolute character of acceleration}

An intriguing issue in general relativity is the nature of 
accelerating or rotating motions. Let us consider two bodies in
space-time, A and B, that describe arbitrary motions. We assume,
for simplicity, that these motions take place 
along a locally straight line. Let us consider the situation in
which (i) A is at rest, with B moving relatively to A, and (ii)
B is at rest, with A moving relatively to B. Are these two
physical situations equivalent? If both situations are equivalent,
we can assert that acceleration is relative. If not,
we conclude that the acceleration of a physical system is absolute.

This issue has been addressed by Mashhoon \cite{Mashhoon1}, who 
investigated the relativity of rotation. He concluded that 
gravitomagnetic effects are an absolute measure of the rotation of
a massive body, and therefore that rotation has an absolute 
character. Mashhoon considers a rotating astronomical body, the
Earth for instance, and the heavens are described to a certain
approximation by a spherical shell. It is observed that the two
situations, the Earth considered at rest, at the center of the
rotating heavens, and the Earth rotating at the center of the
static heavens, produce physically inequivalent effects on 
test particles on the surface of the Earth.

We will conclude that acceleration is absolute by means of a 
totally different analysis. We consider the following two 
situations: (i) an observer linearly accelerated with respect to 
a static black hole, and (ii) a black hole linearly accelerated 
with respect to an observer at rest. In
both situations we assume that the observer is somehow able to 
measure the gravitational radiation across an open spherical 
surface $S_r$, produced by the accelerated motion. 
The surface $S_r$ is established at a fixed distance $r$ from the
center of the black hole, where the observer is located. The 
first situation described above has been investigated in Ref. 
\cite{Maluf5}. 

Before we proceed, let us note that the concept of ``static" 
black hole makes sense provided we establish the usual asymptotic 
boundary conditions $g_{\mu\nu}\cong \eta_{\mu\nu}
+h_{\mu\nu}(1/r)$ in the limit $r\rightarrow \infty$. 
Moreover the absence of a Rindler horizon
ensures the stationary character of the black hole. In Ref. 
\cite{Maluf5} it was considered a moving observer in the
presence of a static black hole. If the observer moves at constant
proper velocity and is sufficiently far from the black hole, 
then the principle of relativity guarantees that
the observer might be considered at rest in the presence of the
moving black hole. We note that in this case the usual tetrad 
field boundary conditions for the black hole, 
$e_{a\mu}\cong \eta_{a\mu}+(1/2) h_{a\mu}(1/r)$, are not 
satisfied. The tetrad field for a
Schwarzschild black hole in isotropic coordinates, that is boosted 
in the $x$ direction, reads \cite{Maluf5}

$$e^a\,_\mu(t,x,y,z)=\pmatrix{\gamma M&-\gamma vN&0&0\cr
-\gamma vM&\gamma N&0&0\cr
0&0&N&0\cr
0&0&0&N\cr}\,,$$
where

$$M^2={{(1-m/2\rho)^2}\over {(1+m/2\rho)^2}}\,,$$

$$N^2=(1+m/2\rho)^4\,,$$
and $x=\rho\sin\theta\cos\phi$, $y=\rho\sin\theta\sin\phi$ and
$z=\rho \cos\theta$.
The variable $\rho$ is related to the usual radial coordinate $r$
by $r=\rho(1+m/2\rho)^2$. In the asymptotic limit 
$\rho \rightarrow \infty$ we have $e_{(0)1}\rightarrow -v\gamma$ 
and $e_{(1)0}\rightarrow -v\gamma$, which are not in agreement with
the ordinary boundary conditions 
$e_{a\mu}\cong \eta_{a\mu}+(1/2) h_{a\mu}(1/r)$.

By considering an observer acellerated in the $x$
direction, in the presence of a static Schwarzschild black hole, 
it has been shown that the gravitational radiation in the 
accelerated frame, across a surface $S_r$, is given by \cite{Maluf5}

\begin{equation}
\Phi^{(0)}_g= 
{1\over {4\pi}} \int_{S_r}
d\theta d\phi\,(\dot{\gamma}\,m\sin\theta)= 
{1\over {4\pi}} \int_{S_r}
d\theta d\phi\,(mv\dot{v}\gamma^3 \sin\theta)\,.
\label{46}
\end{equation}
The total gravitational radiation measured  in the accelerated
frame is

\begin{equation}
\Phi^{(0)}_g = \dot{P}^{(0)} = mv\dot{v}\gamma^3\,.
\label{47}
\end{equation}
We note that the expression above is exact, in contrast to Eq. (44).

The situation in which the black hole is accelerated with respect 
to observers approximately at rest is the subject of section 
4 above. In the context of the present analysis, the 
gravitational radiation across the same surface $S_r$ is given by
expression (46) plus correction terms. Taking into account Eq. 
(42), we see that up to first order in $\varepsilon$ we have

\begin{equation}
\Phi^{(0)}_g \cong
{1\over {4\pi}} \int_{S_r}
d\theta d\phi\,v\dot{v}\gamma^3 \sin\theta ( m+
Ar\cos\theta)\,.
\label{48}
\end{equation}
The expression above is expected to hold in the frame where the
four-velocity of the observers is given by Eq. (34). However at 
the radial distance $r$ such observers are assumed to be 
approximately at rest because Eq. (34) differs
from Eq. (30) by $O(\varepsilon)$ terms. Thus the 
$O(\varepsilon)$ term in Eq. (48) might be questionable. However,
as long as we depart from conditions (31,32), Eq. (48) will differ
from Eq. (46) by several additional terms (according to Eq. (38)). 
Therefore we conclude that the
detection of gravitational radiation in the two situations described
above will be different. The measured values of the radiation can be 
used, in principle, to decide whether the black hole is static or 
accelerated.

\section{Final Remarks}

The analysis of the gravitational radiation emitted by accelerated
particles in the framework of the C metric space-time
has been previously addressed by, among others, Kinnersley 
and Walker \cite{KW}, Farhoosh and Zimmerman \cite{FZ}, Bonnor 
\cite{Bonnor} and Tomimatsu \cite{AT}. Since the C metric is a 
vacuum solution of Einstein's field equations and describe the
exterior space-time of an accelerated particle of mass $m$, the
radiation in question arises from a monopole term, very much like 
the similar phenomenon that takes place in electromagnetism. We 
recall that the analysis of the possible types of gravitational
energy transfer have been investigated by Bondi \cite{Bondi}, who
concluded that in general relativity gravitational energy can 
be transferred by means of inductive or radiative transfer, in 
similirarity to electromagnetism.

In this article we have likewise considered the C metric space-time
and analyzed the issue of gravitational energy transfer by means of
a radiative process. Since the C metric is a vacuum solution of 
Einstein's field equations, the only available form of energy is
gravitational energy, and thus we can only have gravitational
energy transfer. 

We have constructed a set of tetrad fields that establish
the four-velocity of observers that are approximately at rest in
space-time, in the limiting case determined by conditions (31) and
(32), which ensure that the observer is at great distance from the
Schwarzschild and Rindler horizons.
Under these conditions we evaluated the total gravitational
radiation emitted by the accelerated black hole, which turns 
out to be the simple expression given by Eq. (44). Finally we 
concluded that the accelerated motion in space-time is not 
relative. In this respect we 
note that the deformation of the event horizon of a 
Schwarzschild black hole, as determined by $r_S$ presented in 
section 2, constitutes one further distinctive feature of the
accelerated system. The physical properties of the horizon would,
in principle, allow us to characterize a static or accelerated
black hole. 

\bigskip
\noindent {\bf Acknowledgment}\par
\noindent This work was supported in part by the Brazilian Agency
CNPQ.\par

\end{document}